\documentclass[sigconf,authorversion,nonacm]{acmart}

\usepackage{comment}
\usepackage{graphicx}
\usepackage{caption}
\usepackage{subcaption}
\usepackage{url}
\usepackage{xurl}
\usepackage{hyperref}
\usepackage{listings}
\usepackage{float}
\usepackage{soul}
\usepackage[T1]{fontenc}
\usepackage{lmodern}

\DeclareUrlCommand{\tturl}{\urlstyle{tt}}
\DeclareUrlCommand{\bftturl}{}
\DeclareUrlCommand{\bfurl}{}

\AtBeginDocument{
  \providecommand\BibTeX{{
    \normalfont B\kern-0.5em{\scshape i\kern-0.25em b}\kern-0.8em\TeX}}}

\begin{document}

\title{Web Archives for Verifying Attribution in Twitter Screenshots}

\author{Tarannum Zaki}
\affiliation{
  \institution{Old Dominion University}
  \city{Norfolk}
  \state{Virginia} 
  \country{USA}
}
\email{tzaki001@odu.edu}

\author{Michael L. Nelson}
\affiliation{
  \institution{Old Dominion University}
  \city{Norfolk}
  \state{Virginia} 
  \country{USA}
}
\email{mln@cs.odu.edu}

\author{Michele C. Weigle}
\affiliation{
  \institution{Old Dominion University}
  \city{Norfolk}
  \state{Virginia}
  \country{USA}
}
\email{mweigle@cs.odu.edu}

\renewcommand{\shortauthors}{Zaki, Nelson, Weigle}

\begin{abstract}
Screenshots of social media posts are a common approach for information sharing. Unfortunately, before sharing a screenshot, users rarely verify whether the attribution of the post is fake or real. There are numerous legitimate reasons to share screenshots. However, sharing screenshots of social media posts is also a vector for mis-/disinformation spread on social media. We are exploring methods to verify the attribution of a social media post shown in a screenshot, using resources found on the live web and in web archives. We focus on the use of web archives, since the attribution of non-deleted posts can be relatively easily verified using the live web. We show how information from a Twitter screenshot (Twitter handle, timestamp, and tweet text) can be extracted and used for locating potential archived tweets in the Internet Archive's Wayback Machine. We evaluate our method on a dataset of 1,571 single tweet screenshots.
\end{abstract}

\keywords{Twitter/X, misinformation, disinformation, screenshot, web archives}

\maketitle

\section{Introduction}
\begin{figure*}
    \includegraphics[width=\linewidth]{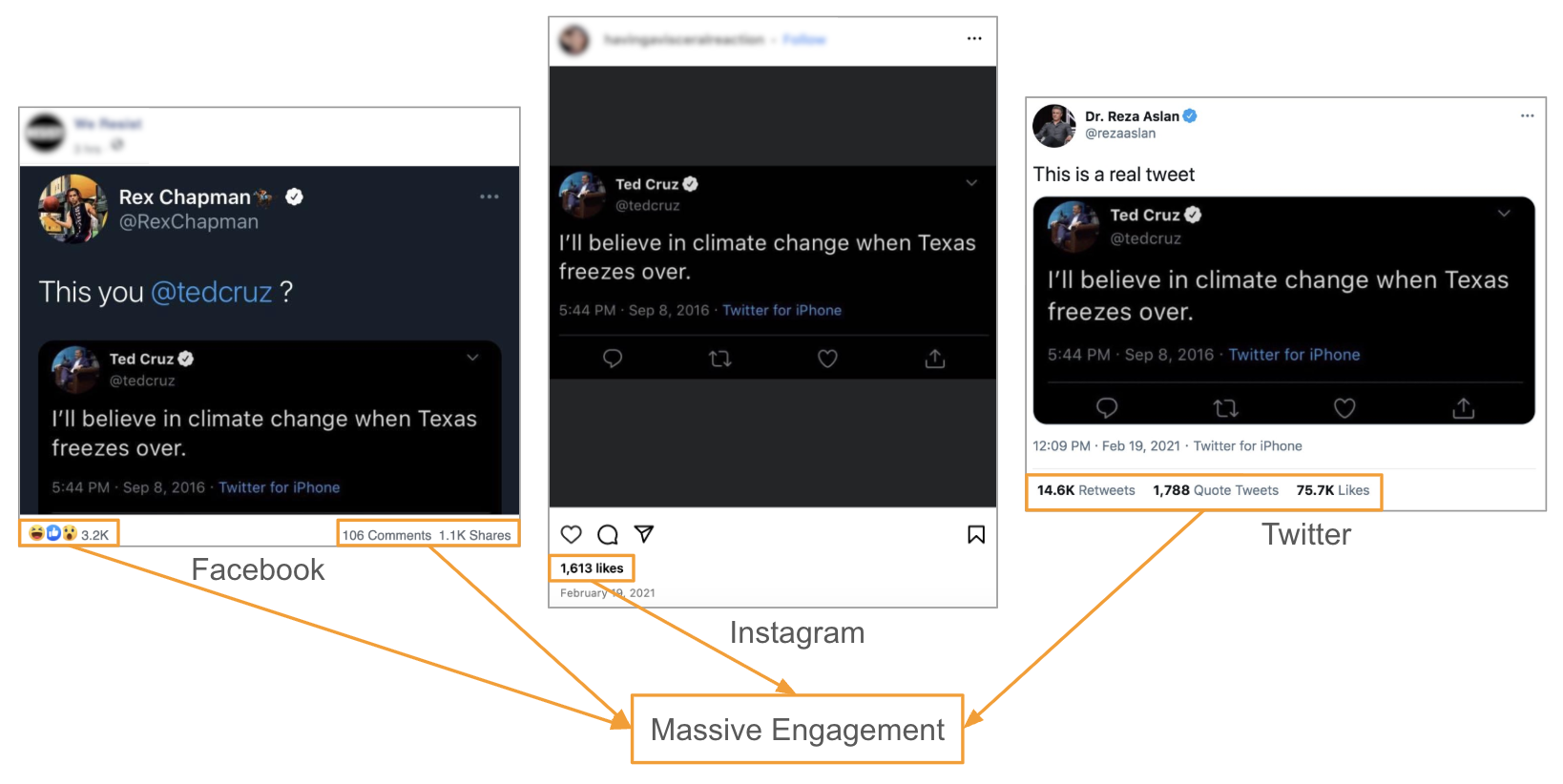}
    \Description{...}
    \caption{Multiple people sharing screenshot of a fake tweet multiple times on different social media platforms made to appear to be by @tedcruz shared on Twitter. The Facebook post was shared over a thousand times, the Instagram post was liked over a thousand times, and the Twitter post was retweeted over 14 thousand times.}
    \label{mas-eng}
\end{figure*}

Twitter\footnote{Twitter is now known as X, and ``tweets'' are now known as ``posts.'' However, we will continue to use the terms ``Twitter'', ``tweets'', etc.} is a social media platform popular among users for sharing information about news, current events, and politics \cite{kwak2010twitter}. As a result, Twitter has become a target for spreading mis-/disinformation. Sharing screenshots of faked tweets has become a common way to mislead social media users. The screenshot of a fake tweet may be shared by multiple people, multiple times, and on different social media platforms. For example, Figure \ref{mas-eng} shows multiple people sharing a screenshot of the same tweet\footnote{For privacy purposes, we blur the Twitter handle of non-public figures.} (which was \emph{not} posted by Ted Cruz) multiple times on Instagram, Twitter, and Facebook \cite{Reuters2021Ted}. Massive engagement with a fake tweet may result in spreading mis-/disinformation. Therefore, it is important to verify a tweet's \emph{attribution}, whether the content of a screenshot was really posted by the alleged author, before sharing.

\begin{figure*}
\centering
  \begin{subfigure}[b]{0.6\textwidth}
  \centering
    \includegraphics[width=1\linewidth]{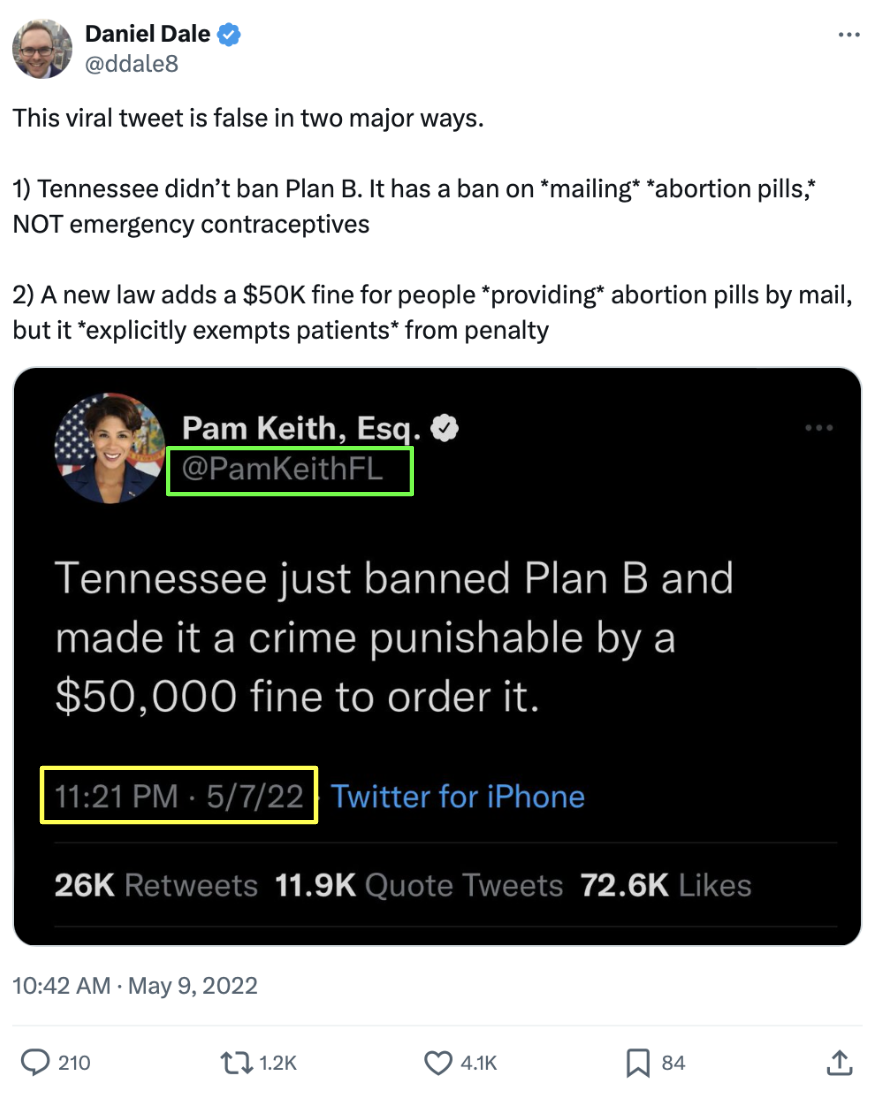}
    \Description{...}
    \caption{A deleted tweet posted by @PamKeithFL.}
    \label{deleted-tweet-pam}
  \end{subfigure}
  \begin{subfigure}[b]{0.7\textwidth}
  \centering
    \includegraphics[width=1\linewidth]{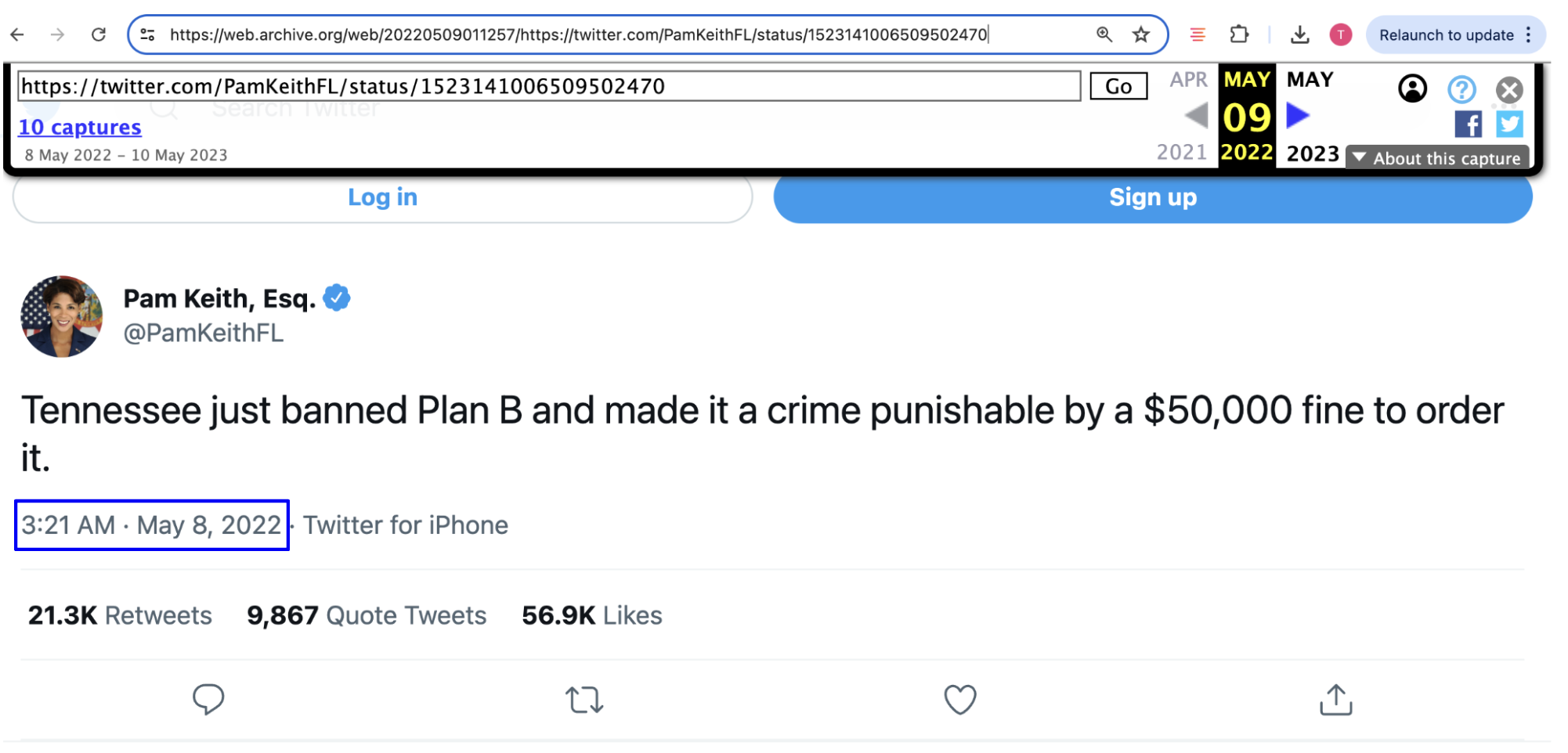}
    \Description{...}
    \caption{Archived version of a deleted tweet posted by @PamKeithFL (\url{https://web.archive.org/web/20220509011257/https://twitter.com/PamKeithFL/status/1523141006509502470}).}
    \label{archived-tweet-pam}
  \end{subfigure}
\caption{Screenshot of a deleted tweet and its archived version.}
\label{pam-example}
\end{figure*}

\begin{figure}
    \includegraphics[width=\linewidth]{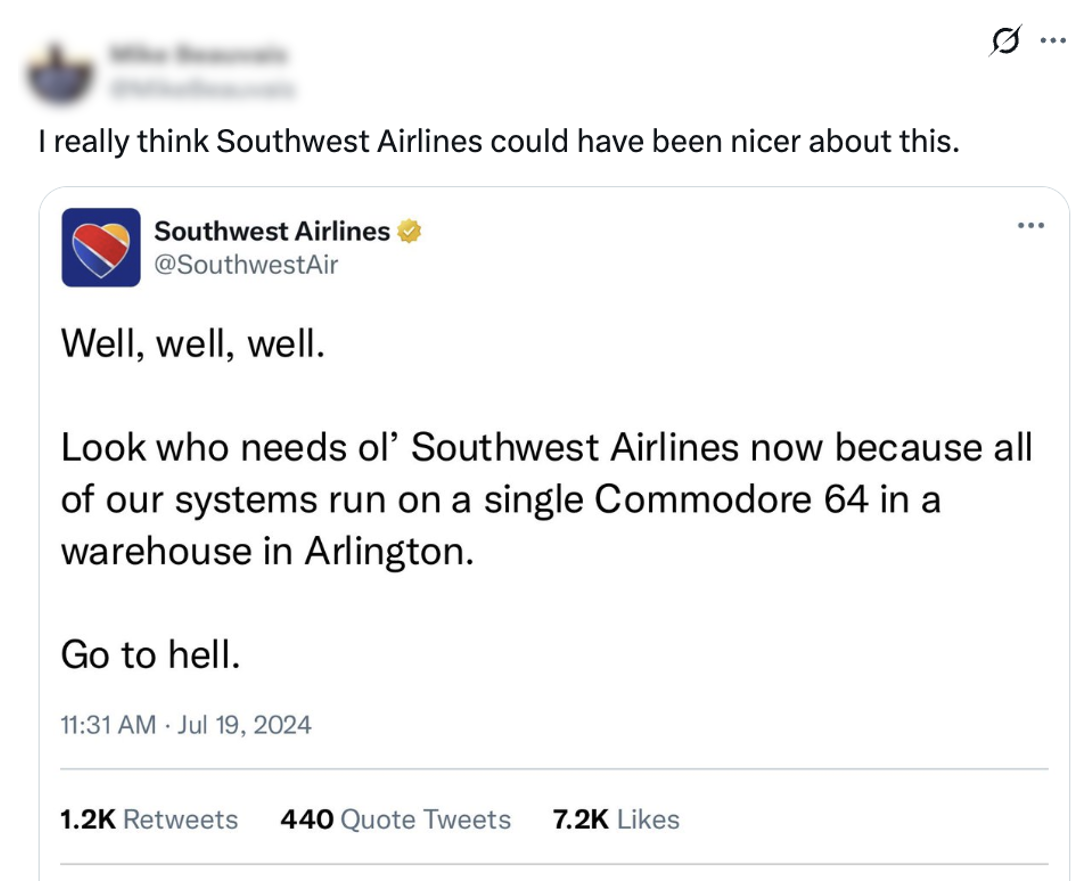}
    \Description{...}
    \caption{Because this humorous fake tweet never happened, it is not on the live web or web archives.} 
    \label{fake-not-found}
\end{figure}

Social media users may take screenshots of posts  to keep them as evidence in case a post (or the entire account) is later deleted. For example, Figure \ref{pam-example} shows that US politician Pam Keith posted a tweet about pregnancy-preventing emergency medication, but later deleted it \cite{Dale2022}. However, an archived version exists in the Internet Archive's Wayback Machine (Figure \ref{archived-tweet-pam}). 
Other reasons to use screenshots  include for humor, as in Figure \ref{fake-not-found}, preventing further engagement with a controversial post, or for cross-platform sharing (e.g., sharing a Twitter post on Facebook). Social media platforms, with the goal of keeping users' attention on their platform, have an incentive to make it difficult for users to share content from other platforms. When users post screenshots as a way around this, they are creating hidden links to another platform. Part of our work involves surfacing these links.

\begin{figure*}
\centering
  \begin{subfigure}{0.75\textwidth}
  \centering
    \includegraphics[width=1\linewidth]{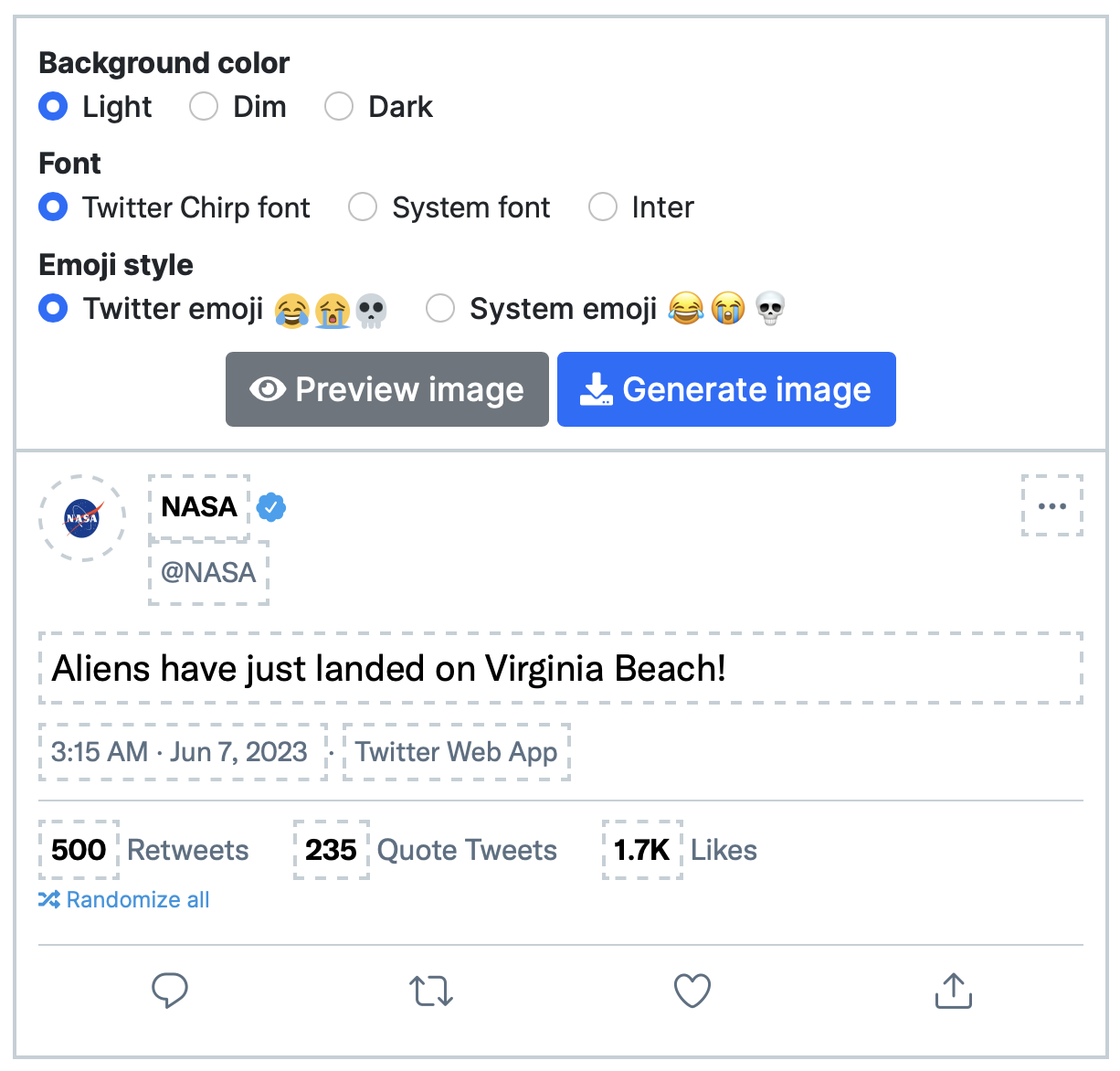}
    \Description{...}
    \caption{Editable attributes to create a fake tweet made to appear to be by @NASA.}
    \label{create-fake-tweet}
  \end{subfigure}
  \vspace{4ex} 
  \begin{subfigure}{0.75\textwidth}
  \centering
    \includegraphics[width=1\linewidth]{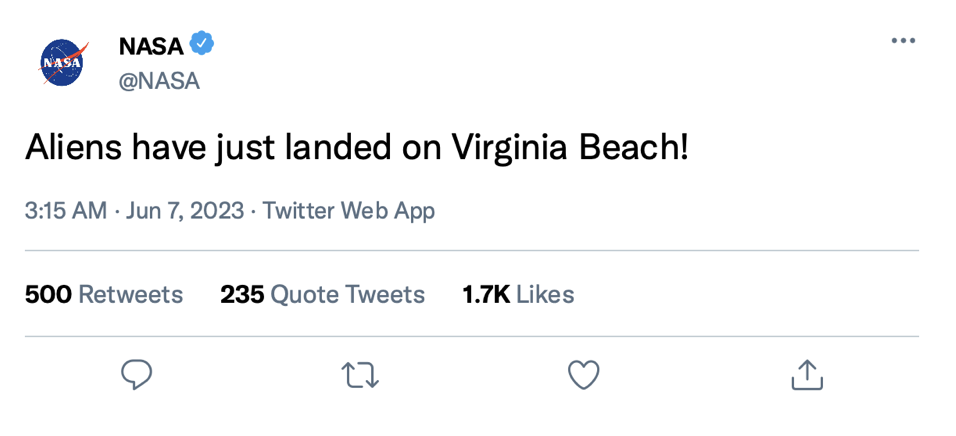}
    \vspace*{-0.8cm}
    \Description{...}
    \caption{A fake tweet from @NASA.}
    \label{fake-tweet}
  \end{subfigure}
\caption{An example fake tweet created using Tweetgen.}
\label{tweetgen}
\end{figure*}

One issue in verifying the attribution of a tweet in a screenshot is that online tools, such as Tweetgen, can be used to easily generate images of fake tweets. These tools also allow users to set like and retweet counts to fake engagement metrics. Figure \ref{tweetgen} shows how Tweetgen\footnote{https://www.tweetgen.com/} provides an interface of editable attributes which can be easily formatted to create a fake tweet. However, there are no automated tools currently available to detect such fake tweets.

\begin{figure*}
    \includegraphics[width=\linewidth]{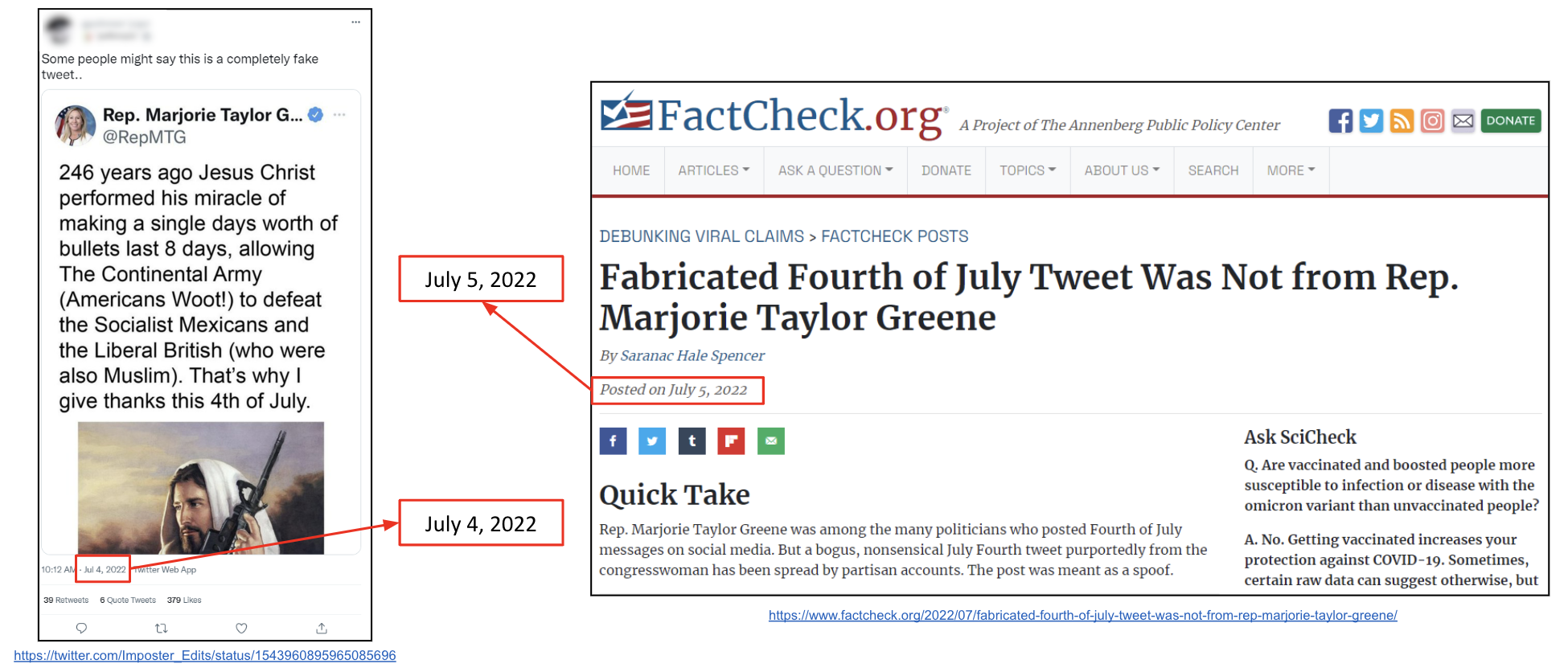}
    \Description{...}
    \caption{Using tweet text to verify the attribution of the tweet in the screenshot using a fact-checking website.}
    \label{factcheck-image}
\end{figure*}

There are manual methods for verifying a tweet's attribution; the simplest is to search for the tweet text in a search engine, such as Google. If the tweet is real, this search may return a link to the tweet on the live web. In addition, fact-checking websites, like Snopes\footnote{https://www.snopes.com/}, Reuters\footnote{https://www.reuters.com/fact-check/}, and FactCheck.org\footnote{https://www.factcheck.org/}, allow users to check the veracity of information. For example, Figure \ref{factcheck-image} shows that FactCheck.org \cite{Spencer2022} verified that the attribution of a screenshotted tweet to Rep. Marjorie Taylor Greene regarding the 4th of July was false (with the appropriate cultural context, it is clearly political satire). 

\begin{figure*}
\centering
  \begin{subfigure}[b]{0.9\textwidth}
  \centering
    \includegraphics[width=1\linewidth]{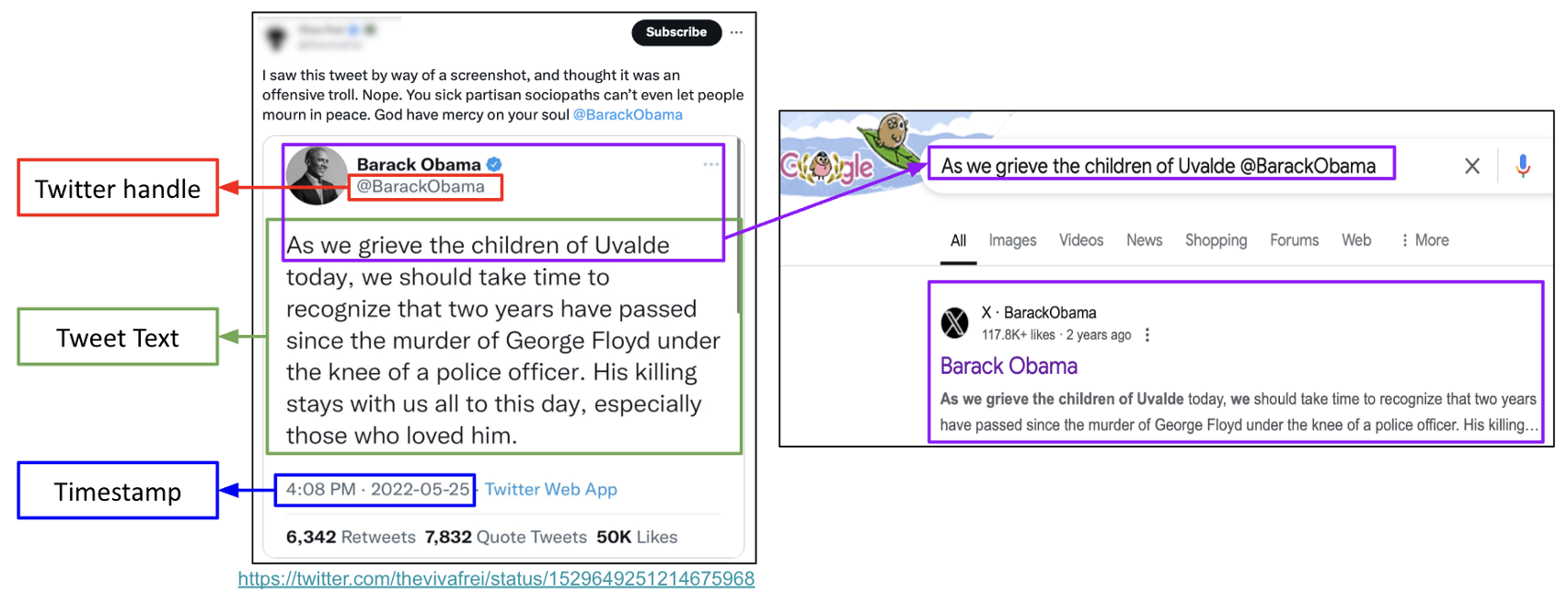}
    \Description{...}
    \caption*{}
    \label{}
  \end{subfigure}
  \vspace{4ex} 
  \begin{subfigure}[b]{0.75\textwidth}
  \centering
    \includegraphics[width=1\linewidth]{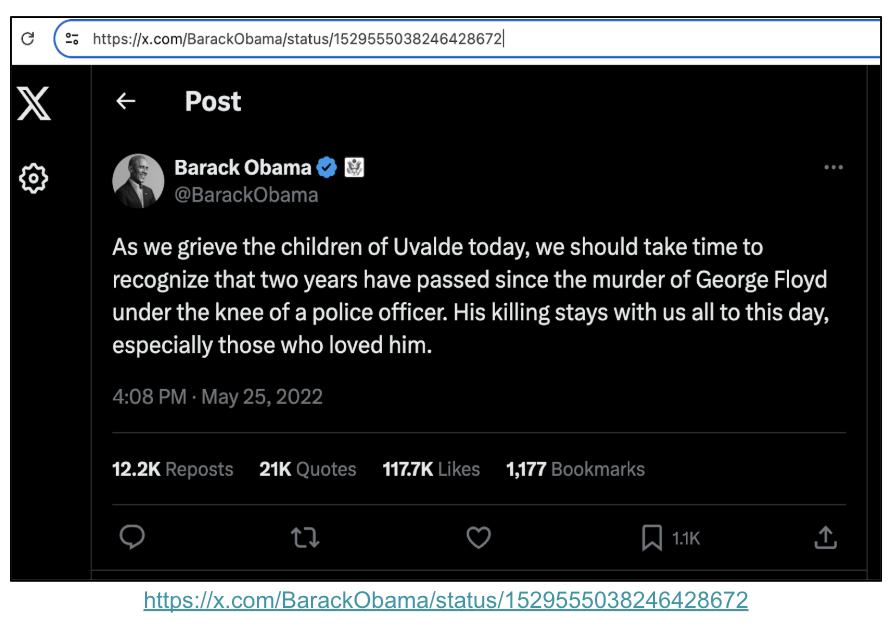}
    \vspace*{-1.5cm}
    \Description{...}
    \caption*{}
    \label{}
  \end{subfigure}
\caption{Using tweet text and Twitter handle to verify the attribution of the tweet screenshot using Google Search.}
\label{tweet-info-google-search}
\end{figure*}

Another useful method is to search web archives, like the Internet Archive's Wayback Machine\footnote{https://web.archive.org/}, especially for deleted posts or accounts (see Figure \ref{pam-example}). Searching for a tweet in a web archive typically requires knowledge of the tweet's URL.  But when we only have a screenshot of a tweet (without its URL), a user would need the author's Twitter handle and tweet timestamp to be able to construct a URL prefix that could be queried using the Wayback Machine's CDX API\footnote{https://archive.org/developers/wayback-cdx-server.html}. 

In this paper, we focus on the use of web archives because the process of searching for a live tweet appearing in a screenshot  using services such as search engines and fact-checking sites is quite straightforward. But, there are cases when we cannot find the alleged tweet on the live web. In these cases, we are unable to determine whether the tweet was faked or was actually posted by the alleged author and later deleted. We rely on web archives to help verify the attribution of screenshots of deleted tweets. Archived deleted tweets serve as evidence that a tweet was really posted by the alleged author, which we cannot prove using live web services. 

Our main goal is to aid in reducing mis-/disinformation spread by providing automated methods to make it easier to verify the attribution of a social media post shared as a screenshot. In this paper, we focus on single tweet screenshots and on using web archives to verify attribution of a tweet.

It is important to note that verifying the validity of the \emph{attribution} of a tweet is different than verifying the validity of the \emph{content} (e.g., ``the sun rises in the west''). This difference is demonstrated in Figure \ref{pam-example}. We count this as a real, or verified, attribution since the tweet was found in the Wayback Machine (Figure \ref{archived-tweet-pam}), but as noted in the tweet in Figure \ref{deleted-tweet-pam}, the content of what the author posted was not true. 

As a part of this overall goal, we first discuss how a human can use the Twitter handle, timestamp, and tweet text from a screenshot of a tweet to search for the alleged tweet on the live web and in web archives. Next, to automate this process we need to be able to extract the Twitter handle, timestamp, and tweet text from a tweet screenshot. For live tweets, one could pay to obtain this information  using the Twitter API or by web scraping the Twitter HTML page. However, our work concerns getting this information from a tweet screenshot, so we provide and evaluate methods for extracting these elements from images. Lastly, we demonstrate how the Wayback Machine's CDX API can be used to verify whether the screenshot tweet was really posted by the alleged author. 

\section{Background}
\begin{figure*}
\centering
  \begin{subfigure}[b]{0.4\linewidth}
  \centering
    \includegraphics[width=\linewidth]{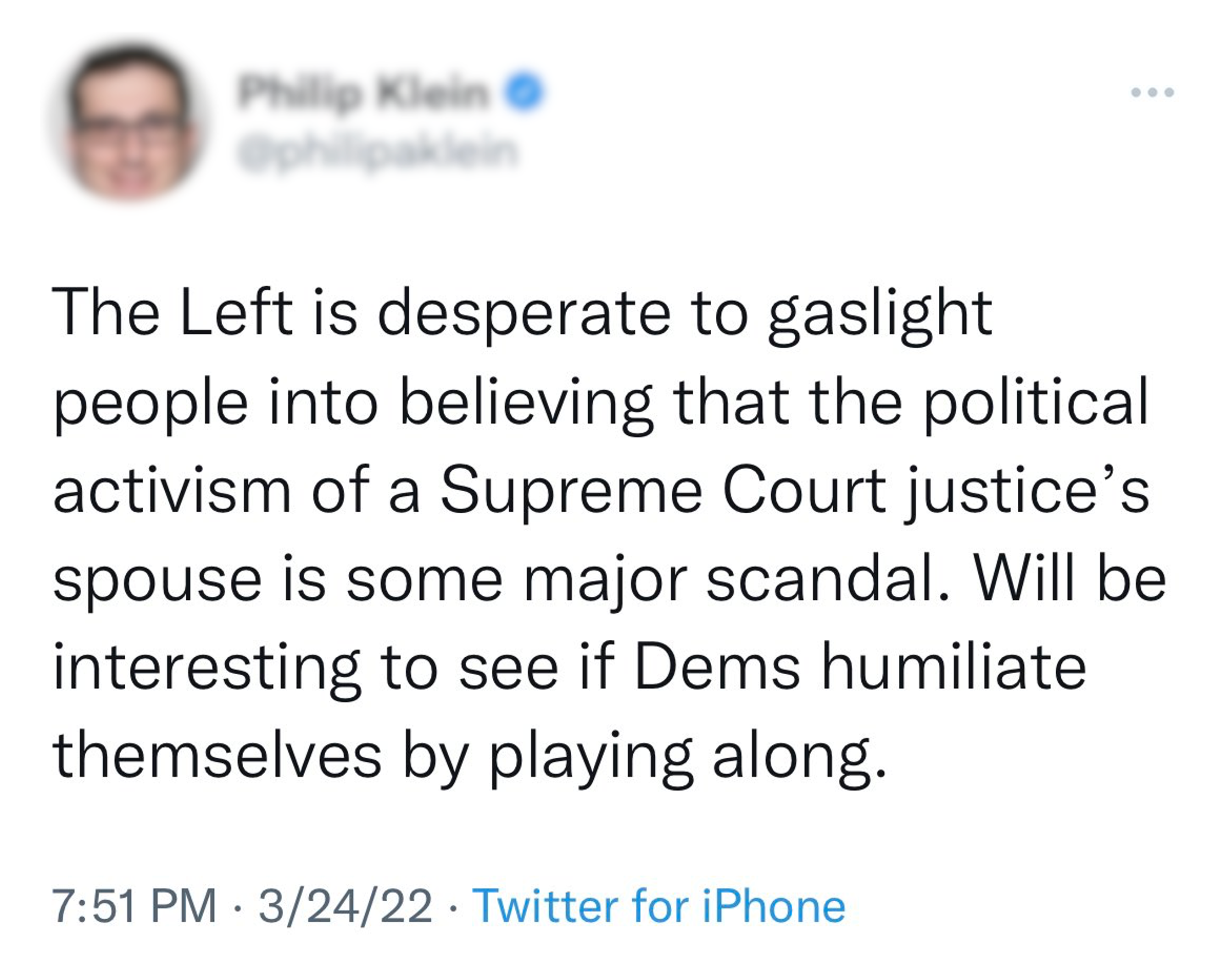}
    \Description{...}
    \caption{Single tweet}
    \label{single-tweet}
  \end{subfigure}
  \hspace{2ex}
  \vspace{2ex}
  \begin{subfigure}[b]{0.45\linewidth}
  \centering
    \includegraphics[width=\linewidth]{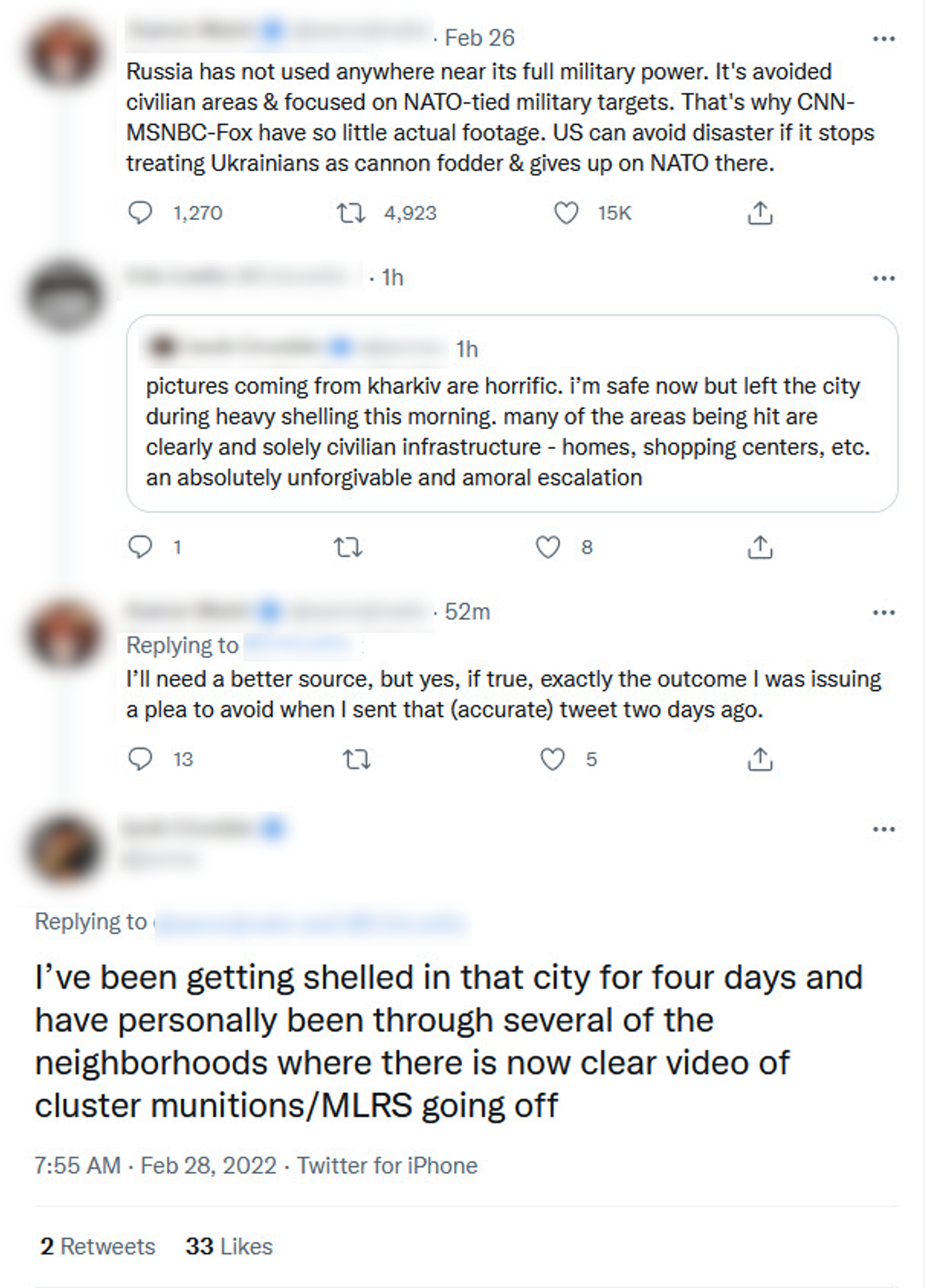}
    \Description{...}
    \caption{Twitter thread}
    \label{thread-tweet}
  \end{subfigure}
  \begin{subfigure}[b]{0.75\linewidth}
  \centering
    \includegraphics[width=\linewidth]{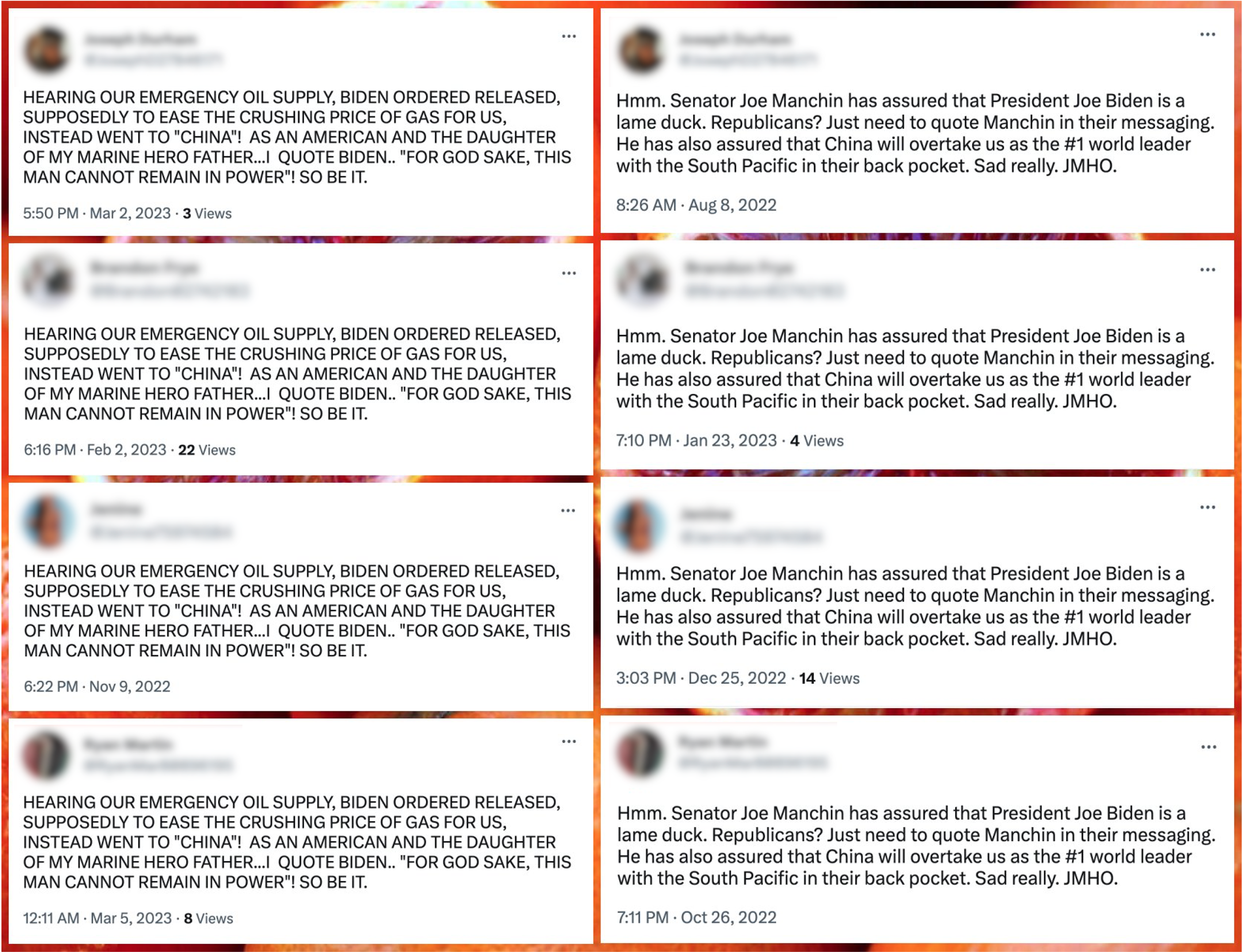}
    \Description{...}
    \caption{Concatenated tweets}
    \label{concat-tweet}
  \end{subfigure}
\caption{Different types of Twitter screenshots.}
\end{figure*}

Our research concerns working with screenshot images of tweets, which can be in different forms. Broadly, screenshots of tweets can be classified into three types: single tweet images, multi-tweet images, and others. We define single tweet images as having a single post, which may contain the post's author and the timestamp (Figure \ref{single-tweet}). We consider multi-tweet images as those having multiple posts which may contain multiple authors and timestamps, such as quote tweet and threads (Figure \ref{thread-tweet}). The ``others'' category consists of tweets that we are unable to classify based on the above-mentioned definitions. An example is a concatenated tweet (Figure \ref{concat-tweet}), which consists of multiple single tweets stitched together. In this paper, we work with single tweet images only.

Our objective is to find evidence on the live web or in the web archives that would help verify the attribution of the alleged tweet. One way this can be done is to use the tweet text to query Twitter's search function, search engines, and fact-checking websites on the live web. Note that deleted tweets would disappear from Twitter's search function, and search engines, though if embedded, those tweets would persist in the context of the embedding page \cite{jones-blog-2021}. Apart from individual fact-checking websites, Google Fact Check Explorer\footnote{https://toolbox.google.com/factcheck/explorer/search} 
specifically provides search results of claim reviews by different fact-checking websites. In addition to these, Politwoops\footnote{https://projects.propublica.org/politwoops/} 
Politwoops tracked the deleted tweets of public officials, though as of 2023 it is no longer operational due to Twitter's API change. Fact-checking sites are typically focused on the truth value of the content, but we focus on truth value of the attribution.

Another method to verify the attribution of a tweet is to search web archives, such as the Wayback Machine, using the tweet's URL, since many web archives are indexed by URL. The URL of a tweet,\footnote{The structure of the URL for a post on X is the same as for a tweet, except with \texttt{x.com} replacing \texttt{twitter.com}.} such as \tturl{https://twitter.com/}\bftturl{PamKeithFL}/status/\bftturl{1523141006509502470}, contains a Twitter handle (PamKeithFL) and a tweet ID (1523141006509502470). Each tweet ID is a unique identifier that encodes its creation date, using the Snowflake service \cite{snowflake}. We can use the TweetedAt tool \cite{Siddique2019}
\footnote{\url{https://oduwsdl.github.io/tweetedat/}} 
to determine the tweet post time using its tweet ID.
\begin{figure*}
    \includegraphics[width=\linewidth]{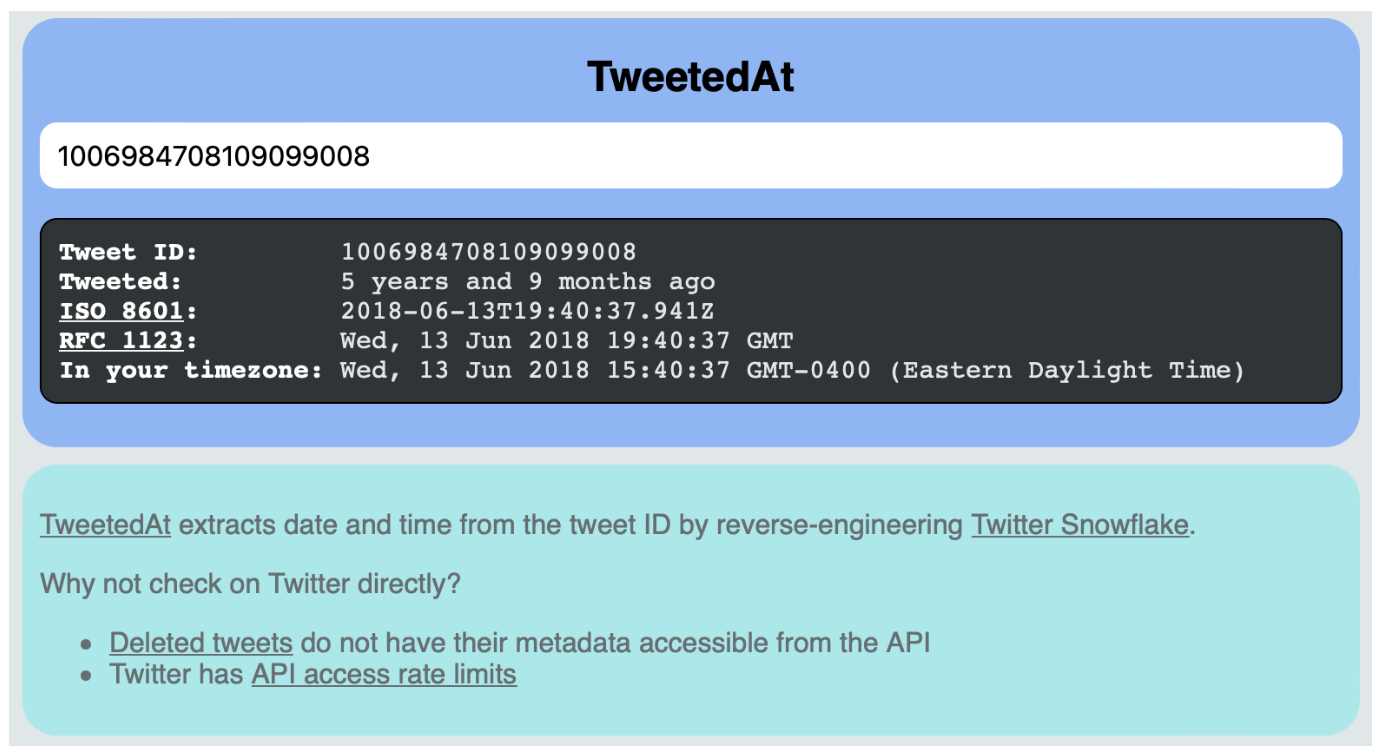}
    \Description{...}
    \caption{Determining tweet creation timestamp using TweetedAt (\url{https://oduwsdl.github.io/tweetedat/\#1006984708109099008})}
    \label{tweetedat}
\end{figure*}

However, we cannot directly predict a tweet ID or its URL from a screenshot. Tweet ID is a unique identifier for each tweet and cannot be predicted in advance. 
This is different from other URLs such as news articles (e.g., \url{https://www.cnn.com/sport/live-news/paris-olympics-news-2024-08-08/index.html}), which could potentially be predicted in advance. As a result, we cannot predict a tweet URL from a screenshot to search in the web archives. 

\section{Related Work}
The spread of mis-/disinformation on social media is a widespread issue, because it is easy to create and spread false information but a daunting task to detect the falsehood. Aïmeur, Amri, and Bassard \cite{Aimeur2023} identified  that obtaining a fake news dataset representing a real-world scenario is a crucial challenge in mis-/disinformation research. Shu et al. \cite{Shu2020} presented FakeNewsNet, a multidimensional data repository that contains metadata of news-related tweets such as user profiles, user engagement, posts, etc. Golbeck et al. \cite{Golbeck2018} presented a dataset of fake news and satire stories and further performed a thematic content analysis on the dataset. Our research to detect mis-/disinformation involves determining attribution from visual data containing text -- screenshot images of tweets rather than direct textual data such as metadata, URLs, etc. A part of our dataset also represents a real-world scenario of Twitter screenshot examples, both real and fake, shared on social media \cite{zaki-blog-2022, zaki-dataset}. We also collected a dataset of screenshots \cite{Farris2024, dataset-ashlyn24} to categorize to which social media platform a screenshot belongs. The dataset contains screenshots from Facebook, Instagram, Twitter, and Truth Social. We augment our collected shared tweets with Twitter screenshots from this dataset. Our dataset is not  only limited to smartphone screenshots but also consists of web screenshots in both light and dark modes.

Optical character recognition (OCR) is a widely used technique for working with images that contain text. Kamal et al. \cite{Kamal2023} used OCR via the Google Cloud Vision API on news images to extract text. They later cross-checked the text with \emph{The New York Times} database for fact-checking. Chiatti et al. \cite{Chiatti2018} used OpenCV image processing and Tesseract OCR modules on smartphone screenshots of daily activities and evaluated the quality of the extracted text. They showed that further improvements are possible by fine-tuning image pre-processing parameters. Kumar et al. \cite{Kumar2020} used the Tesseract OCR module on smartphone screenshots to extract keywords and combined the keywords with image features to generate relevant tags for a screenshot.
We use Microsoft Azure's AI Vision \cite{MSazureAPI} to extract text from our dataset of screenshots of tweets.

Geeng et al. \cite{Geeng2020} performed a behavioral study to investigate how people interact with misinformation on Facebook and Twitter. They found that users find it time consuming and mentally overwhelming to further investigate evidence for misinformation while browsing their social feeds. The authors suggested that using some means of automation like alerting users  through fact-checking would help users to simplify their efforts of investigation. Our research focuses on establishing a framework that would automatically provide some evidence to verify attribution of screenshots using the services of live web and web archives for detecting misinformation on social media. In previous work, we developed a module \cite{Bradford2022} using services of the live web, Google Search, and fact-checking sites (Reuters.com, Snopes.com, and Politwoops) to detect fake tweets. The module queries the tweet text in  Google Search to find relevant articles from fact-checking sites and scrapes the truth rating. The module also searches for any evidence from Google Search, such as a Twitter URL. We will be integrating this module in our system as part of searching the live web to verify the attribution of the tweet in Twitter screenshots. Buntain and Golbeck \cite{Buntain2017} developed an automated mechanism to classify whether popular Twitter threads belong to real/fake news stories. They used two datasets for credibility assessment, a crowd sourced dataset and a journalistic dataset, and found that crowd sourced models perform better than journalistic models for detecting fake stories in Twitter. There may be cases where credibility is related to attribution, but we focus only on attribution.

Parikh et al. \cite{Parikh2019} proposed a tweet verification framework to validate Twitter screenshots. The authors used Microsoft Azure’s Computer Vision to extract metadata from screenshots with which to  query the Twitter API. Finding the tweet on the live web guarantees attribution, but the failure to find a tweet does not guarantee non-attribution (individual tweets and entire accounts can be deleted). We do not use Twitter API because of its restricted access and high cost. Moreover, we will not find evidence for deleted tweets by querying the Twitter DB, so we use web archives to search evidence for deleted tweets. Shao et al. \cite{Shao2016} designed Hoaxy, a platform that collects, detects, and analyzes misinformation based on fact-checking efforts. The authors collected tweet instances sharing URLs of fake news and fact-checking information of those fake news and found that misinformation spreads largely than fact-checking content on social media. Starbird et al. \cite{Starbird2014} investigated the propagation of rumors and fake news through social media. They explored the rumors that spread on Twitter after the 2013 Boston Marathon bombing and found that correction to  misinformation lagged behind compared to the propagation of misinformation. Hodges, Chaiet, and Gupta \cite{Hodges2021} performed a forensic analysis on how Twitter screenshots propagate on social media leading to misinformation spread. They introduced SMOC BRISQUEt, a method that analyzes propagation history of a shared image by measuring the level of compression artifacts present on the image. Much of this research lead to detecting propagation of mis-/disinformation on social media.

Web archives contribute to misinformation research by being a useful resource to verify whether some content is real or fake. Weigle \cite{Weigle2023} provided an overview on how web archives can be used for investigating webpage changes, studying archived social media, and validating claims about past statements in disinformation research. Chalkiadakis et al. \cite{Chalkiadakis2021} queried the Wayback Machine's CDX API to analyze the state of fake news websites by retrieving an index of the timestamps for a particular website. Alonso et al. \cite{Alonso2017} implemented a search engine that features a `Wayback Machine' functionality. The authors experimented with a dataset of URLs collected from viral tweets. A user can query the search engine for a particular topic and further can retrieve information about what was said about the topic in the past. Our work involves the use of the Wayback Machine's CDX API, but to query the tweet URLs, we first construct the URLs using information from Twitter screenshots. 

Prior studies typically work with Twitter URLs whereas we work with Twitter screenshots. While previous works show assessing credibility of tweet content, we focus on attribution of the tweet. Unlike other works that use Twitter API to find a tweet, we use web archives because of the API's limited accessibility to research. Some research works analyze propagation history of a shared image whereas our research highlights surfacing the hidden links created through cross-platform sharing of screenshot images.

\section{Verifying the Attribution in Tweet Screenshots}
When provided with an image of a tweet, our goal is to find evidence on the live web or in web archives of the original tweet. If we are able to find some evidence, then we have verified the attribution. We try a variety of methods to do this using the following information from a Twitter screenshot: Twitter handle, timestamp, and tweet text (Figure \ref{tweet-info-google-search}). The timestamp shown in the screenshot is the time of the tweet in the timezone of the person who took the screenshot. It was 4:08 PM on May 25, 2022, but we do not know exactly when Obama posted the tweet -- it may not have been the same timezone. We can identify a real attribution using the live web using search engines and a fake attribution using fact-checking sites. We can also find a real attribution in web archives. But, if we are unsuccessful at all of them and no evidence is found, then the likelihood is higher that the post in the screenshot is fake (assessing this probability is the focus of our ongoing work).
 
\subsection{Real attribution, found on the live web}
Figure \ref{tweet-info-google-search} shows how tweet text from a screenshot can be used to backtrack to the original URL using Google Search. Here, a substring (``As we grieve the children of Uvalde'') from the shared screenshot and the Twitter handle (@BarackObama) are used as a Google Search query to find whether anything relevant exists on the live web. The search returns the original Twitter URL of the post by @BarackObama. This confirms that the content of the shared screenshot was  posted by @BarackObama. Searching for tweets using the Twitter API is not feasible because the cost of API access is now prohibitively expensive.\footnote{In February 2023, Twitter discontinued free access to its API \cite{Calma2023}, which once served as an integral part of social media research.} Moreover, the rate limitations of the API restrict data retrieval within a certain time period. A user could search with Twitter's Advanced Search feature, but that option is not available for automated methods.

\subsection{Fake attribution, found in fact-checking sites}
Figure \ref{factcheck-image} shows how the tweet text from a screenshot can be used to verify the attribution of the tweet in the screenshot using a fact-checking website. A screenshot of a tweet with fake attribution to @RepMTG was shared on Twitter on July 4, 2022. We input the tweet text in the fact-checking website FactCheck.org's search field to find whether any relevant fact-checking article exists. The search provides a relevant article that verifies on July 5, 2022 that the content was not posted by @RepMTG \cite{Spencer2022}. They checked the author's official Twitter feed as well as Politwoops. The verification process of FactCheck.org indicates that the content of the shared screenshot was not posted by @RepMTG.

\subsection{Real attribution, found in web archives}
To search for an archived tweet, we need to know the tweet's URL. Since a tweet's URL is not usually present in a screenshot, we need to construct the URL for the tweet with only the information present in the  screenshot. 

Figure \ref{deleted-tweet-pam} shows a screenshot of a tweet that was originally posted by @PamKeithFL but later deleted. From this, we can see the Twitter handle (@PamKeithFL) and the timestamp (11:21 PM May 07, 2022). The timestamp shown in the screenshot is the time of the tweet in the timezone of the person who took the screenshot, but we do not know exactly when the author posted the tweet -- it may not have been the same timezone. 

\begin{figure*}
    \includegraphics[width=\linewidth]{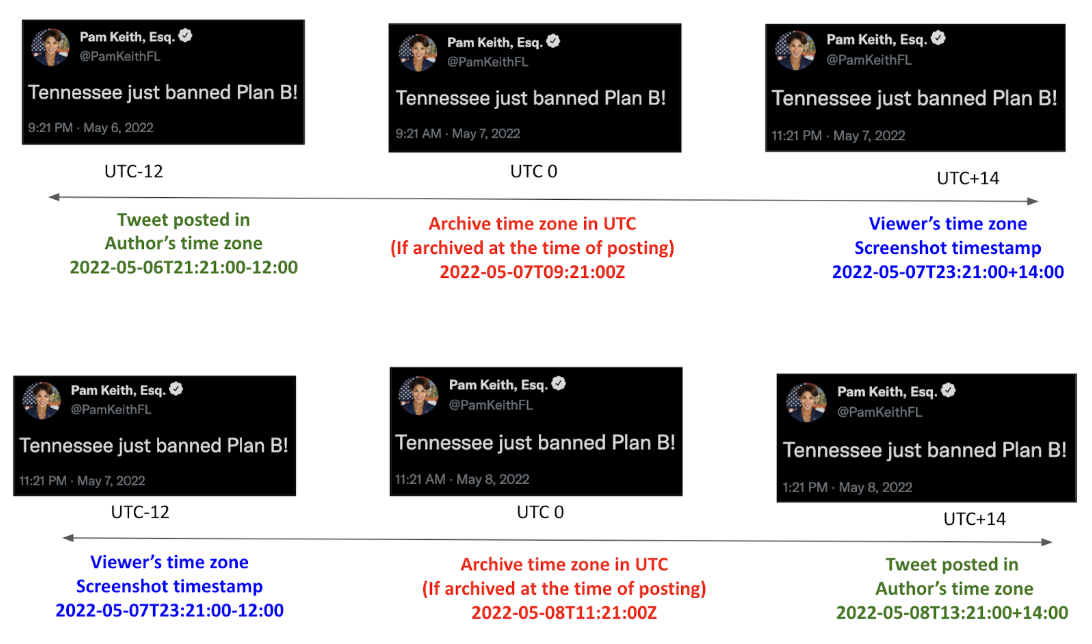}
    \Description{...}
    \caption{Determining a reasonable estimate of the earliest archival time of a tweet to search in web archives.}
    \label{time52h}
\end{figure*}

To search in web archives, we need to construct a reasonable estimate of the earliest time the tweet could have been archived, based on UTC, which is the timezone used by the Wayback Machine. Figure \ref{time52h} demonstrates some possible scenarios.
We can determine a reasonable left-hand boundary \texttt{(from=)} for searching using the CDX API. This is because the timestamp on the screenshot (annotated with yellow box in Figure \ref{deleted-tweet-pam}) is in the timezone of the person who took the screenshot. We do not know or cannot predict the timezone of the person who took the screenshot. For example, the user taking the screenshot may be in Line Island (world's furthest forward time zone, UTC+14) and the user tweeting may be in Baker Island (world's furthest behind time zone, UTC-12), or vice versa. This demonstrates that the maximum time difference on Earth is 26 hours. 
Assume that the timestamp on the screenshot is `May 07, 2022 11:21 PM' in the time zone of the viewer who took the screenshot. If the viewer is in the UTC+14 time zone and the tweet author is in the UTC-12 time zone, then the author's timestamp of the tweet would be 26 hours behind the viewer's timestamp. Similarly, if the time zones are swapped between viewer and author, the author's timestamp of the tweet would be 26 hours before the viewer's timestamp. Consequently, the archival time window for the alleged author's tweet could be approximated as ±26 hours from the timestamp in the screenshot. Hence, it would be reasonable to use a left-hand boundary of 26 hours before the screenshot timestamp for searching in the web archive. We do not need to limit the right-hand boundary strictly as posts may be archived at a later time.

Given the user's handle, we can find all of the archived tweets for that user from the Wayback Machine's CDX API with a query for \texttt{https://twitter.com/}\emph{TwitterHandle}\texttt{/status/*}. We can use this constructed URL and the curl command, as shown in Figure \ref{code-snippet}, to access the CDX API. 
\begin{figure*}
\lstset{moredelim=[is][\color{red}]{[*}{*]}}
\Description{...}
\begin{lstlisting}[breaklines, basicstyle=\footnotesize\ttfamily]
curl -s "http://web.archive.org/cdx/search/cdx?url=https://twitter.com/PamKeithFL/status&from=20220506212100&matchType=prefix" | sort -u -k 3 | awk '{print "https://web.archive.org/web/" $2 "/" $3};'

https://web.archive.org/web/20220508183923/https://twitter.com/PamKeithFL/status/1523140142177058816
https://web.archive.org/web/20220508031859/https://twitter.com/PamKeithFL/status/1523140247739117574
https://web.archive.org/web/20220508032025/https://twitter.com/PamKeithFL/status/1523140577818296321
https://web.archive.org/web/20220508032058/https://twitter.com/PamKeithFL/status/1523140760107122689
[*https://web.archive.org/web/20220509011257/https://twitter.com/PamKeithFL/status/1523141006509502470*]
\end{lstlisting}
\caption{Code snippet to retrieve URLs from the Wayback Machine using the CDX API.}
\label{code-snippet}
\end{figure*}
To obtain @PamKeithFL's status URLs from the web archive, we use the \texttt{matchType=prefix} parameter, which returns results that match the author's status URL prefix \\
(\texttt{https://twitter.com/PamKeithFL/status}). The timestamp field is set to the corresponding starting timestamp \\ (\texttt{from=20220506212100}, 26 hours before the timestamp shown in Figure \ref{pam-example}) in order to retrieve any relevant archived URLs starting from that timestamp. The archived version of the URL for the deleted tweet is colored in red in the list of retrieved URLs in Figure \ref{code-snippet}. Figure \ref{archived-tweet-pam} shows the retrieved URL, the archived version of the deleted tweet by @PamKeithFL. We can also see that the archived tweet's timestamp is `May 08, 2022 3:21 AM' which is 4 hours after the screenshot timestamp, falling within the ±26 hour time window.

If we find that the tweet in the screenshot has been archived, this proves that the tweet once existed. Finding an archived tweet allows us to prove its existence.
However, we note that not all tweets are archived, so \emph{not} finding a copy does not disprove its existence.

\subsection{Fake attribution, but not found anywhere}
Figure \ref{fake-not-found} shows an example of a shared screenshot of a tweet that was not posted by the alleged author. We did not find any evidence on the live web or in the web archives to support that this tweet was really posted by @SouthwestAir. With cultural context, we know this is a satirical tweet that went viral due to Microsoft's global outage in July 2024.

\section{Methods for Extracting Information from Screenshots}
We have seen how the Twitter handle, timestamp, and tweet text from screenshot images can be helpful to verify the attribution of the tweet in screenshots using the live web and web archives. Here, we introduce methods for extracting this information from screenshot images and evaluate the methods on our dataset. We have evaluated our methods for extracting this information from the OCR-converted text using 4,504 single tweet screenshot images. We first created a dataset of 400 screenshots of tweets (272 real, 69 fake, and 59 undetermined) that were shared on Twitter \cite{zaki-blog-2022, zaki-dataset}. Among these, 260 are single tweet images. We then augmented this dataset with   4,244 single tweet screenshot images \cite{Farris2024, dataset-ashlyn24}. These additional images are screenshots of tweets, not necessarily screenshots that were shared in tweets.

Given a screenshot, we use the Microsoft Azure AI Vision API \cite{MSazureAPI} for OCR and then attempt to identify the Twitter handle, timestamp, and tweet text from the OCR output. OCR errors, emojis, and incomplete screenshots (e.g., timestamps cropped out) are some of the errors we can encounter. 

\begin{figure*}
    \includegraphics[width=\linewidth]{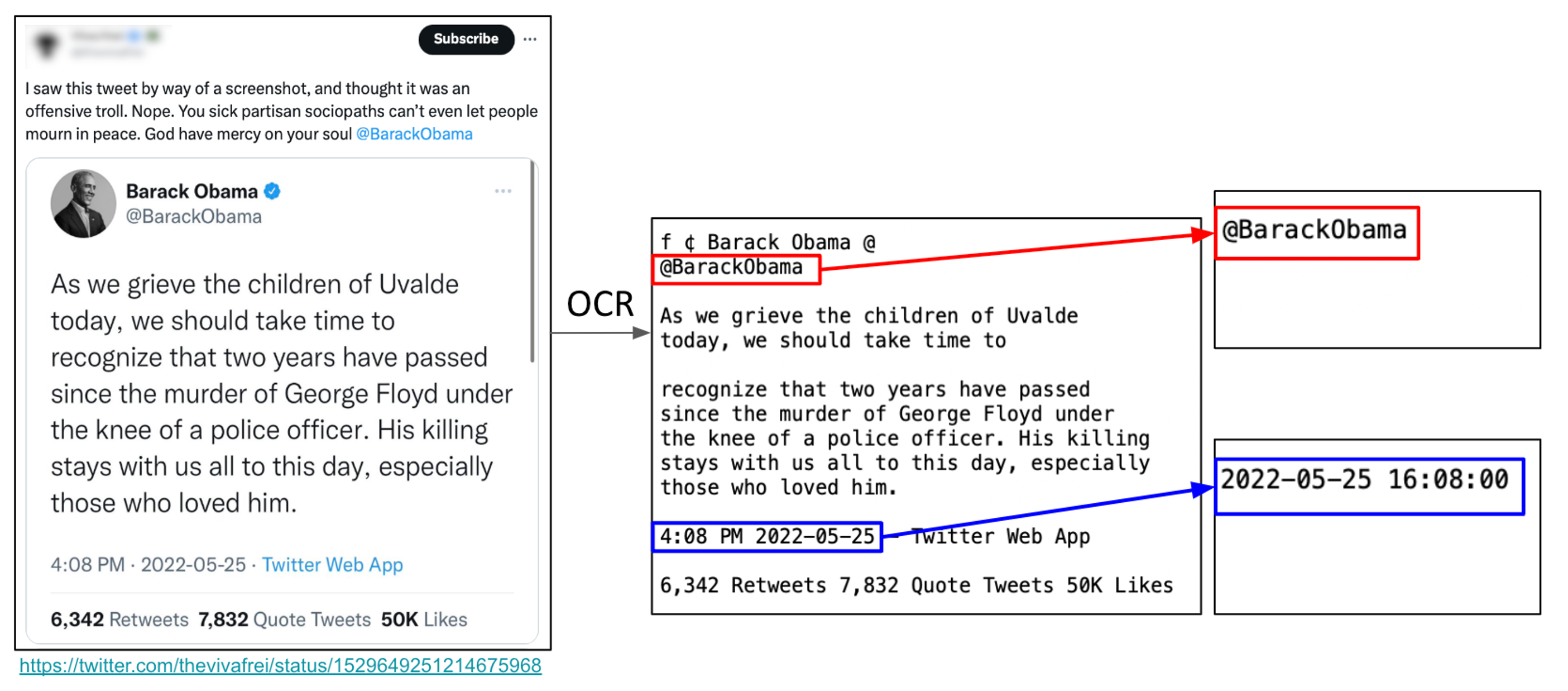}
    \Description{...}
    \caption{Extracting Twitter handle and timestamp from screenshot image.}
    \label{timestamp_handle}
\end{figure*}

\begin{figure}
         \centering
         \includegraphics[width=\linewidth]{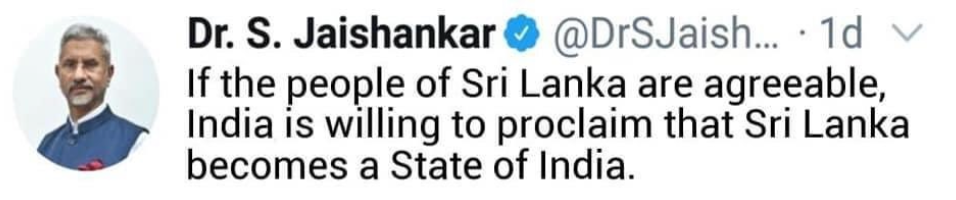}
         \Description{...}
        \caption{An incomplete Twitter handle (@DrSJaish...) and relative timestamp (1d).}
        \label{timestamp-exception}
\end{figure}

We use the Python datefinder module \cite{datefinder} to identify and extract potential timestamps, which are formatted differently depending on the Twitter client's preferences. The datefinder module extracts date type strings from text. Figure \ref{timestamp_handle} demonstrates that the timestamp from the OCR converted output (4:08 PM 2022-05-25) is extracted as 2022-05-25 16:08:00 (annotated with blue colored box). But, the datefinder module may also interpret other numerical digits as timestamps. 
In order to deal with this discrepancy, we use additional logic in the date format. We require at least 6 characters (including the separator) and at least 4 digits to define a fully described date for this logic. We also checked whether the extracted date on the timestamp is valid: 2006$\leq$year$\leq$current\_year (Twitter was founded in 2006), 1$\leq$month$\leq$12, and 1$\leq$day$\leq$31. If two valid timestamps are extracted, we consider only the last one, assuming that timestamp in a tweet is usually in the lower part of the tweet image.
However, there are limitations to this method. For example, screenshots that do not contain timestamp in a particular date time format, such as relative timestamps (Figure \ref{timestamp-exception}), will not result in any output.

For extracting the Twitter handle of a post from a screenshot, some aspects need to be considered. A Twitter handle starts with the `@' symbol. There might be multiple Twitter handles in a single tweet. For example, such a case is when other users are mentioned in a tweet. But we want the Twitter handle of the author only. Moreover, the verified check mark may be converted to a `@' symbol by OCR (e.g., OCR output of Figure \ref{timestamp_handle}). To handle these cases, we traverse through each line of the OCR extracted text, considering words that start with `@' that are not only `@', and extracting the first one among such matched words. Figure \ref{timestamp_handle} demonstrates this method and shows only the Twitter handle of the post (@BarackObama) is extracted from the OCR converted output (annotated with red colored box). Similarly, like the timestamp extraction, there are exceptions where it is not possible to extract the complete Twitter handle. For example, Figure \ref{timestamp-exception} shows a screenshot example where the Twitter handle is incomplete - `@DrSJaish...'

We evaluated our method of timestamp and Twitter handle extraction. Table \ref{table-eval-time-handle} shows the accuracy, precision, recall, and F1 score for 4,504 screenshots. For both methods, we get a high accuracy (100\% for timestamp and 91\% for Twitter handle). For screenshots with Twitter handles containing special symbols (e.g., @\_\_tom\_heard99\_ ), we found that the OCR may extract the symbols inaccurately. Thus, the extracted Twitter handle might be inaccurate at times.

\begin{table}
  \caption{Performance evaluation of the methods for extracting timestamp and Twitter handle (n=4,504)}.
  \label{table-eval-time-handle}
  \begin{tabular}{ccccc}
    \toprule
    \textbf{Info Extracted} & \textbf{Accuracy} & \textbf{Precision} & \textbf{Recall} & \textbf{F1 Score} \\
    \midrule
     Timestamp&100\%&100\%&100\%&100\% \\
     Twitter handle&91\%&92\%&99\%&95\% \\
    \bottomrule
  \end{tabular}
\end{table}

Lastly, we extract the tweet text by utilizing the extracted Twitter handle and timestamp from the OCR extracted output. The tweet text is placed in between the Twitter handle and timestamp (Figure \ref{tweet-info-google-search}). Once we extract a Twitter handle and timestamp from a screenshot, we retrieve the OCR extracted text segment that starts below the identified Twitter handle and ends above the identified timestamp. If we are unable to extract a Twitter handle, timestamp, or both, then we consider the entire OCR extracted output of those screenshots for further processing as we cannot retrieve only the tweet text using the above-mentioned process. Among the 4,504 screenshots in our dataset, we were able to retrieve both Twitter handle and timestamp completely for 3,945 images. We also retrieved tweet text for all these images. 

\section{Verifying Attribution of Screenshots using Web Archives}
\begin{figure}
         \centering
        \includegraphics[width=\linewidth]{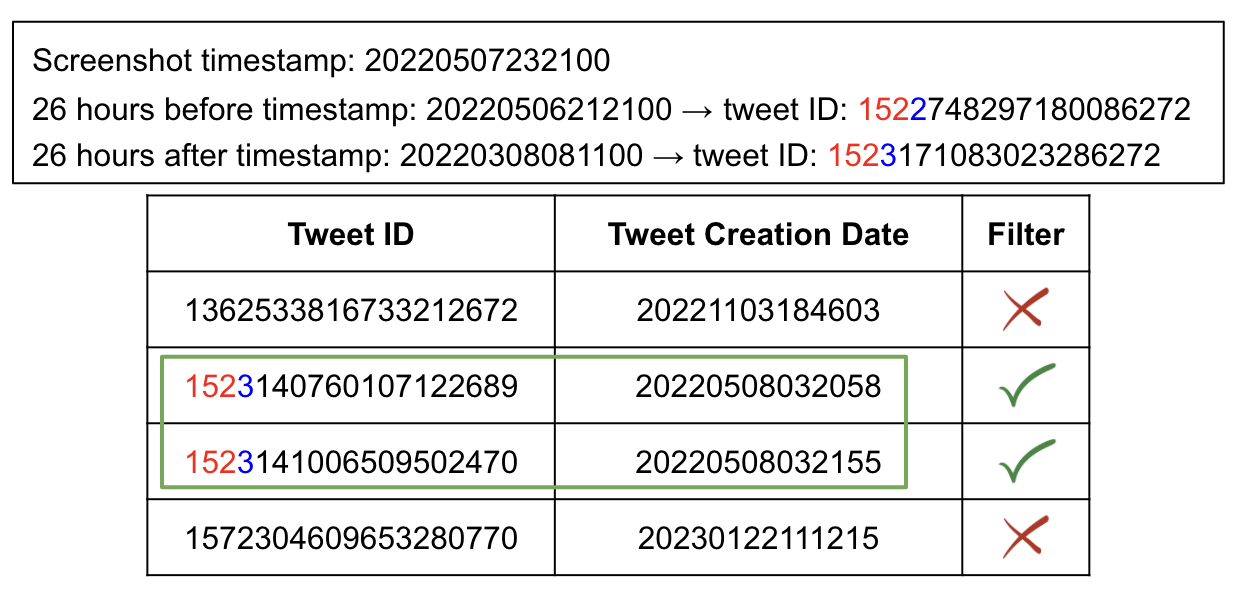}
         \Description{...}
        \caption{Filtering tweet IDs that does not fall within the ±26 hours time window.}
        \label{narrow}
\end{figure}

We have discussed earlier how we use the Twitter handle and left-hand boundary of the screenshot timestamp to search for the screenshot content in the Wayback Machine using the CDX API (Figure \ref{code-snippet}). 
We can further reduce the search space. As explained earlier, each tweet ID encodes its creation date, which we can determine using TweetedAt. 
If we have a target post time, we can reverse the TweetedAt code to determine a range of possible tweet IDs. Though the reconstructed tweet ID from a timestamp will not be exact, it would allow us to filter the request to the CDX API based on the common prefix of those tweet IDs. We can use this to reduce our list of archived tweets to only those that were created within ±26 hours of the screenshot timestamp. Figure \ref{narrow} shows how we select a common Twitter ID prefix of the tweet IDs computed from the left-hand and right-hand boundary timestamps using the reverse TweetedAt to filter the archived tweets within the time window. For example, the two tweet IDs in the green box are the ones we consider for further processing because they fall within the time window. We modify the curl command from Figure \ref{code-snippet2} to reduce the search space of the list of archived tweets as: 
\begin{figure*}
\lstset{breaklines=true, basicstyle=\footnotesize\ttfamily}
\Description{...}
\begin{lstlisting}
curl -s "http://web.archive.org/cdx/search/cdx?url=https://twitter.com/PamKeithFL/status/152[2-3]&from=20220506212100&matchType=prefix" | sort -u -k 3 | awk '{print "https://web.archive.org/web/" $2 "/" $3};'
\end{lstlisting}
\caption{Code snippet of modified curl command to reduce the search space of the retrieved URLs from the Wayback Machine using the CDX API.}
\label{code-snippet2}
\end{figure*}
We use the regex filter \texttt{152[2-3]} parameter to filter the tweet IDs based on the common Twitter ID prefix.
This process greatly reduces the number of archived tweets to examine for a match. For example, the original CDX query shown in Figure \ref{code-snippet} for @PamKeithFL's deleted tweet example returns 16,950 archived tweets. The modified CDX query reduces this  to 378 archived tweets, which we further reduce to 179 by filtering out duplicate archived tweets.

\begin{figure*}
    \includegraphics[width=\linewidth]{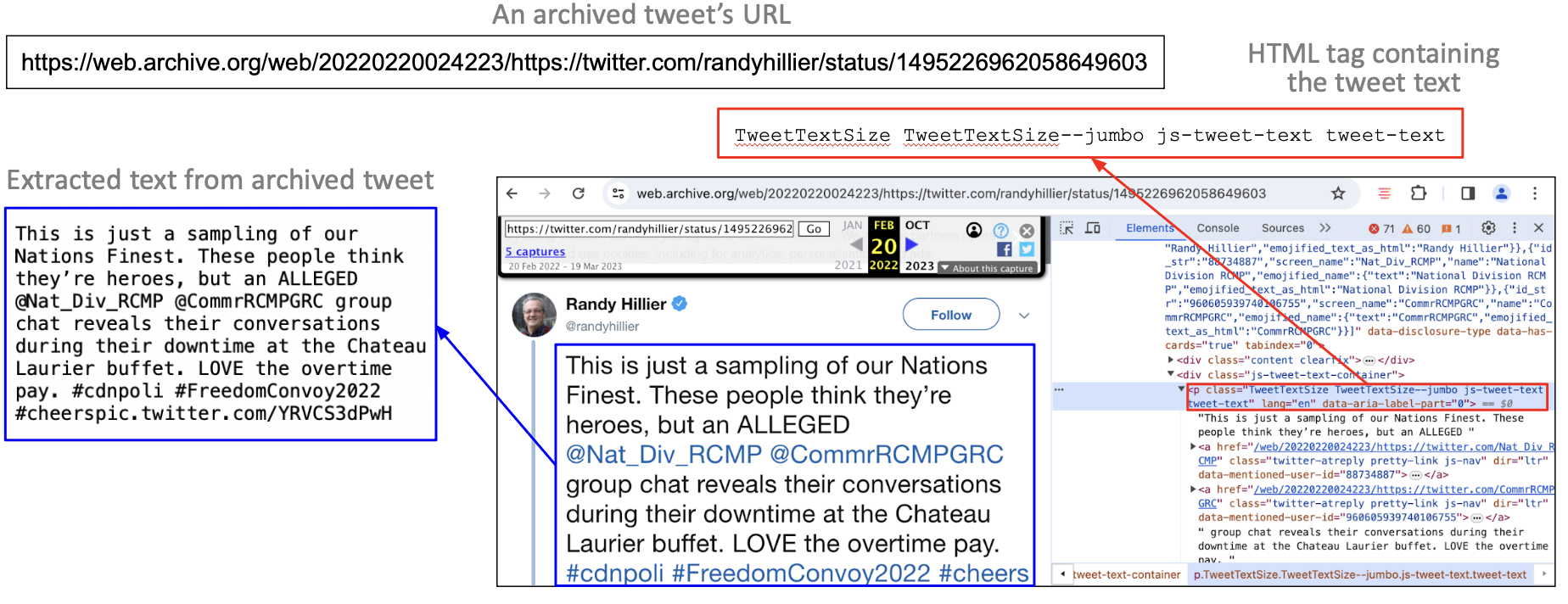}
    \Description{...}
    \caption{Extracting tweet text from archived tweets using BeautifulSoup and Selenium.}
    \label{sel-bs}
\end{figure*}

After obtaining the filtered list of archived tweets, we use Selenium\footnote{https://www.selenium.dev/} 
 and BeautifulSoup\footnote{https://pypi.org/project/beautifulsoup4/}
to load each archived candidate tweet URL and extract the tweet text (Figure \ref{sel-bs}).

\begin{figure*}
\centering
  \begin{subfigure}[b]{0.85\textwidth}
    \includegraphics[width=1\linewidth]{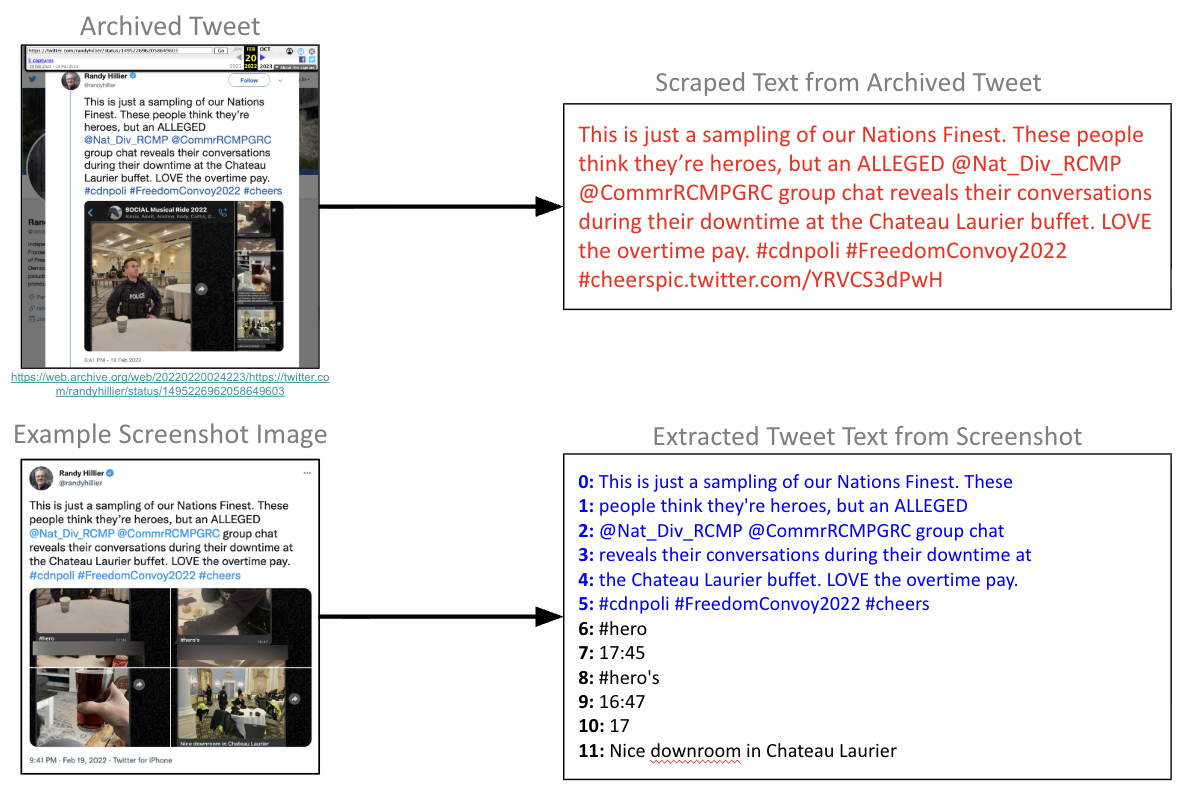}
    \Description{...}
    \caption{}
    \label{tweet_text_extract}
  \end{subfigure}
  \vspace{4ex} 
  \begin{subfigure}[b]{0.85\textwidth}
    \includegraphics[width=1\linewidth]{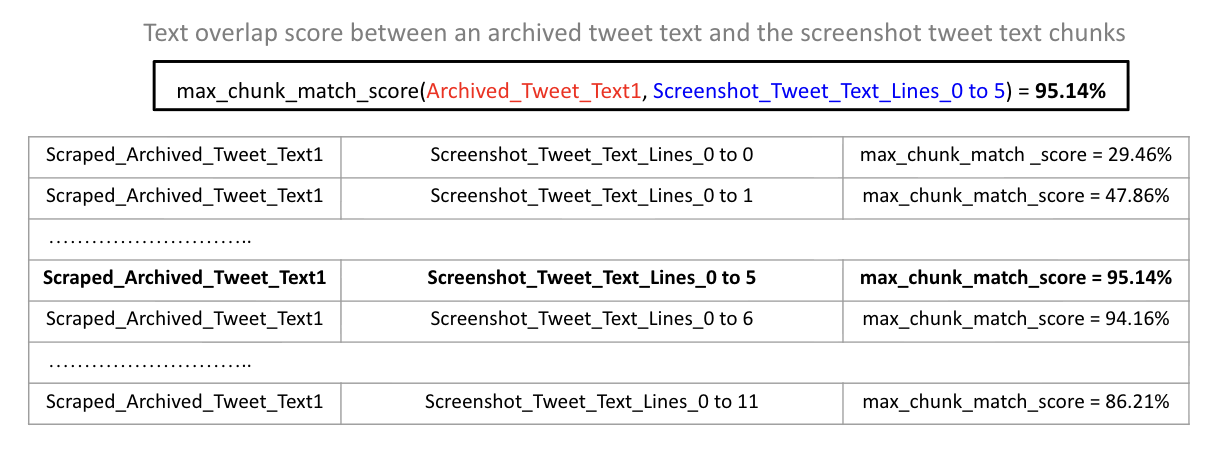}
    \Description{...}
    \caption{}
    \label{text_overlap_score}
  \end{subfigure}
\caption{Computing text overlap score between archived tweet and screenshot tweet text.}
\label{text_overlap}
\end{figure*}

\begin{figure*}
    \includegraphics[width=\linewidth]{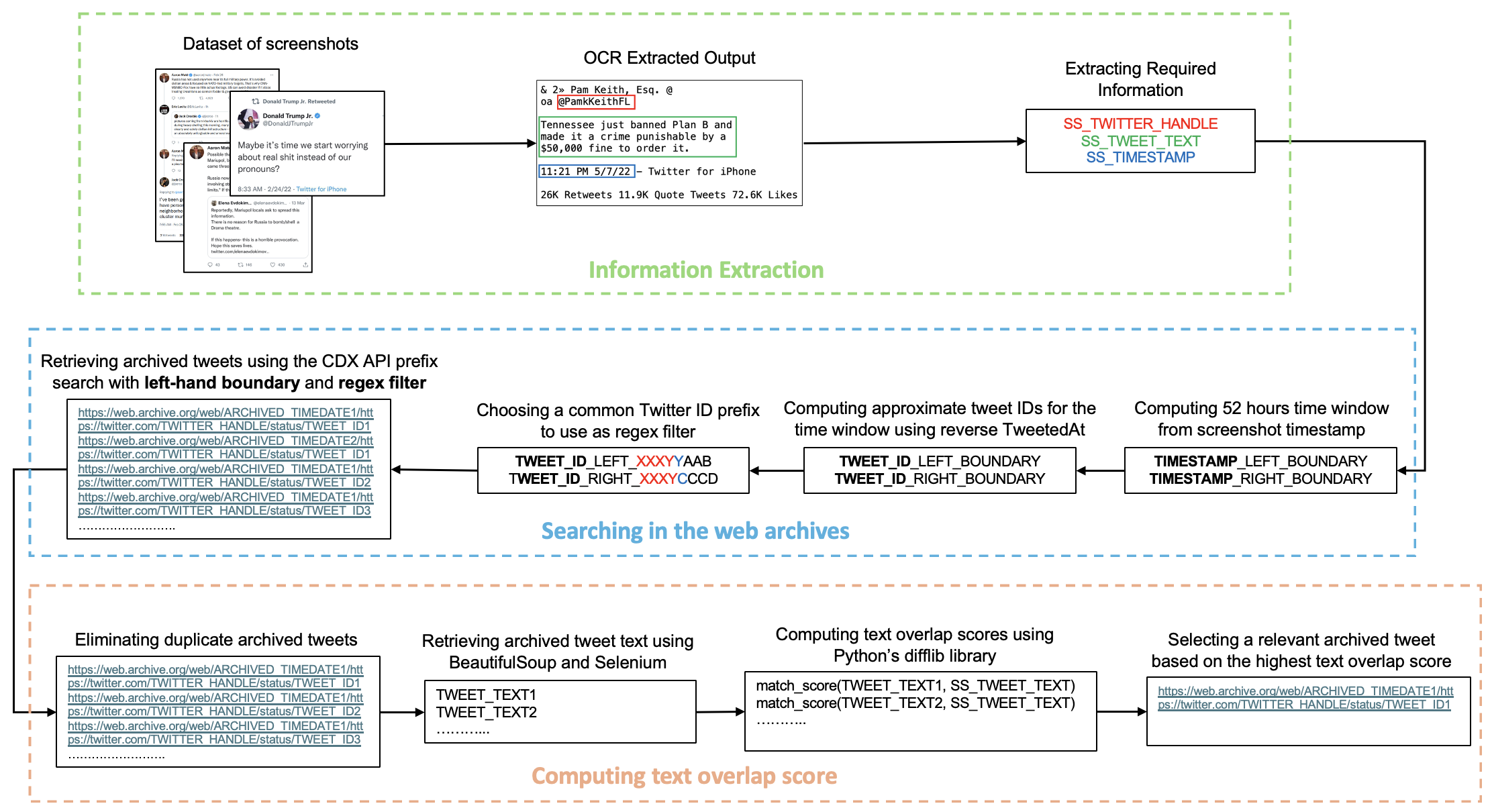}
    \Description{...}
    \caption{Framework for verifying attribution of screenshots using web archives.}
    \label{framework}
\end{figure*}

We then use Python's difflib module to compute a text overlap score based on the longest common subsequence. We first compute the text overlap scores between a candidate tweet's text and chunks of the screenshot's tweet text. Figure \ref{text_overlap_score} shows how we gradually concatenate lines towards the end to compute text overlap scores. We finally chose the maximum score as the one having the most similarity with a specific candidate tweet. We consider an incremental chunk-by-chunk text comparison because OCR extracted text may contain noisy output. Figure \ref{tweet_text_extract} shows that the OCR extracted output also contains text from the attached image which is not part of the tweet text. While comparing with a candidate tweet, such cases can reduce the text overlap score. We further compute text overlap scores between all the candidate archived tweets and the screenshot tweet and consider the one with the maximum score as an evidence of the screenshot tweet posted by the alleged author. Figure \ref{framework} shows a framework for verifying screenshots' attribution using web archives. 

Among 3,945 images in our dataset, we processed 1,571 single tweet screenshots that had unique content. The ground truth for 1,571 tweets are: 69 are archived, 1,497 are not archived, and 5 are non-replayable. Table \ref{table-all-count} shows the total number of archived tweets found using our method for searching in the web archive (n=1,571) at different threshold levels. The threshold represents the lowest text overlap score (80\%) where we found an archived version of a screenshot tweet.

\begin{table}
  \caption{Number of archived tweets found using our method for searching in the web archives (n=1,571).}
  \label{table-all-count}
  \begin{tabular}{ccc}
    \toprule
    \textbf{Our Method} & \textbf{State of Tweets} & \textbf{FP/FN} \\
    \midrule
    \textbf{Threshold 80\%}&Archived: 64&FP: 0, FN: 5 \\
    \textbf{Threshold 90\%}&Archived: 60&FP: 0, FN: 9 \\
    \bottomrule
  \end{tabular}
\end{table} 
We found that at least 80\% text overlap produced the highest F1 (0.96) and had no false positives and minimized false negatives (Table \ref{table-eval}).
\begin{table}
  \caption{ Performance evaluation of highest text overlap scores of the retrieved archived URLs (n=1,571).}
  \label{table-eval}
  \resizebox{8cm}{!}{
  \begin{tabular}{cccc}
    \toprule
    \textbf{Threshold Value} & \textbf{Precision} & \textbf{Recall} & \textbf{F1 Score} \\
    \midrule
    90\%&1.00&0.87&0.93 \\
    \textbf{80\%}&\textbf{1.00}&\textbf{0.93}&\textbf{0.96} \\
    \bottomrule
  \end{tabular}}
\end{table}

We started collecting our dataset in 2022 when Twitter URLs were archived under the domain \texttt{twitter.com}. Since May 17, 2024, the domain of \texttt{twitter.com} officially changed to \texttt{x.com}, and browsers now redirect URLs having \texttt{twitter.com} to \texttt{x.com} \cite{Peters2024}. Moreover, the Wayback Machine's ability to archive archive Twitter/X URLs fluctuates through time \cite{garg2024challenges}. Among the 4,504 screenshots of our collected dataset, 24  were collected after the domain change occurred. Among the 24 screenshots, we found only 2 to be archived, 17 to be not archived, and 5 to be non-replayable. We were not able to verify the attribution of four screenshots using the Wayback Machine's CDX API. This is because an archived URL's presence in the CDX list does not guarantee that we can extract the archived tweet. There might be failure in archiving or replaying, which we have not explored further in this paper. Although the Wayback Machine's ability to archive and replay tweets is variable, we still consider it for our study because it had been a dominant social media platform. Further, the Wayback Machine's CDX API allows us to integrate results from the web archive with other research tools, and this is not available in other web archives.

\section{Conclusions and Future Work}
Screenshots are an easy way to share content on social media. There are many legitimate reasons to share screenshots of social media posts, such as keeping evidence of a deleted post, cross-platform sharing, and humor/satire. But, this also opens the door for mis-/disinformation. Our work focuses on validating the attribution of a screenshot, for example: ``Did this person actually tweet what this screenshot says they did?'' To verify a tweet's attribution in a screenshot, we discussed how the Twitter handle, timestamp, and tweet text can be used to search for content on the live web and in web archives. Next, we introduced methods for extracting these key elements from screenshots. Our evaluation results on 4,504 single tweet screenshot images showed 100\% accuracy for timestamp and 91\% accuracy for Twitter handle extraction. We were able to extract tweet text for 3,945 screenshots among 4,504 based on the existence of both Twitter handle and timestamp. We further demonstrated a process of using the Wayback Machine's CDX API to search in a web archive for screenshot content that would help to verify a tweet's attribution. We emphasized demonstrating the process of searching in web archives because attributing screenshots of deleted tweets is quite challenging using live web services. We performed this process for 1,571 screenshots of unique tweets. We evaluated the text overlap score between the archived and screenshot tweets' text and found that a threshold of 80\% produced the highest F1 score. The collected dataset and the tool we are developing for this research would greatly contribute to the research area of mis-/disinformation spread on social media.

For future work, we plan to introduce methods of extracting information from complex screenshots (e.g., quote tweets, threads, concatenated tweet images). We also plan to extend our study for other social media platforms (e.g., Instagram, Facebook, Truth Social) by first determining which social media platform a screenshot belongs. The user interface (UI) of social media platforms can change over time and we want to explore the impact of different UI generations  (e.g., 2014 Twitter vs. 2024 X). We have worked with the Wayback Machine's CDX API to search for screenshot content in the Internet Archive, and we would further investigate methods to look into other archives including those that do not have CDX APIs. We also intend to use search engines to look for embeds, which persist even if the tweet or account is deleted \cite{jones-blog-2021}. Finally, failing to find existence of a tweet (live web, archived web), we would determine evidence of it being real based on additional factors, such as the alleged author's publishing frequency, the archiving frequency of the alleged author's account, and assessing the reputation signal of authors sharing Twitter screenshots.

\begin{acks}
This work supported in part by the GROW M\&S project (Grant \# 300747-010), funded by the US Department of Education.
\end{acks}

\balance
\bibliographystyle{ACM-Reference-Format}
\bibliography{sample-base}

\end{document}